\pdfoutput=1
\documentclass{appolb}
\usepackage[utf8]{inputenc}
\usepackage{graphicx}
\usepackage{grffile}
\usepackage{amsmath}
\usepackage{subcaption}
\usepackage{comment}

\newcommand{\Nf}{N_{\text{f}}}

\newcommand{\xshift}{\ensuremath{x_{\mathrm{shift}}}}

\begin{document}

\eqsec
\title{Inhomogeneous phases in the 1+1 dimensional Gross-Neveu model at finite number of fermion flavors\thanks{Presented at the  excited QCD conference, Schladming, Austria, 30.01 - 02.02 2019}}

\author{Laurin Pannullo$^a$, Julian Lenz$^b$, Marc Wagner$^a$, Björn Wellegehausen$^b$, Andreas Wipf$^b$\\
\address{
	$^a$Institut für Theoretische Physik, Goethe Universität Frankfurt am Main, \\ 60438 Frankfurt am Main, Germany\\
	$^b$Theoretisch-Physikalisches-Institut, Friedrich-Schiller-Universität Jena, \\ 07743 Jena, Germany}
}

\maketitle

\begin{abstract}
We study the phase diagram of the 1+1 dimensional Gross-Neveu model at finite number of fermion flavors using lattice field theory. Numerical results are presented, which indicate the existence of an inhomogeneous phase, where the chiral condensate is a spatially oscillating function.
\end{abstract}



\section{The Gross-Neveu model and its phase diagram in the limit of infinitely many fermion flavors}

Exploring the phase diagram of QCD using lattice computations is currently restricted to small chemical potential, because of the QCD sign problem (see e.g.\ \cite{Philipsen:2010gj,Aarts:2015tyj} and references therein). There are, however, several QCD-inspired models, e.g.\ the Gross-Neveu (GN) model \cite{Gross:1974jv}, which are technically simpler to treat, and which share certain symmetries with QCD. Studies of such models might, thus, provide insights concerning the phase diagram of strongly interacting matter. A notable feature of the GN model in 1+1 dimensions in the limit of infinitely many fermion flavors is the existence of a so-called inhomogeneous phase, where the chiral order parameter is not a constant, but spatially oscillating \cite{Thies:2003kk,Schnetz:2004vr} (for a review on inhomogeneous condensates and phases see \cite{Buballa:2014tba}). In this work we perform a lattice field theory study of the 1+1 dimensional GN model at \emph{finite number} of fermion flavors $\Nf$, to explore, whether inhomogeneous phases also exist at finite $\Nf$.

The Euclidean action of the GN model is
\begin{align}
	\label{eq:action}
	S = \int \mathrm{d}^2x \, \bigg( \sum_{n=1}^{\Nf} \bar{\psi}_n \Big(\gamma^0(\partial_0 + \mu)  + \gamma^1 \partial_1 \Big) \psi_n - \frac{\lambda}{2 \Nf}\bigg(\sum_{n=1}^{\Nf} \bar{\psi}_n \psi_n\bigg)^2 \bigg) ,
\end{align}
where $\psi$ denotes a fermionic field with $\Nf$ flavors and $\mu$ is the chemical potential. One can get rid of the four-fermion interaction by introducing a scalar field $\sigma$, which leads to the following partition function:

\begin{align}
\label{EQN001}
& Z = \int D\sigma \, \exp{\underbrace{\bigg(-\Nf \bigg(\frac{1}{2 \lambda} \int \mathrm{d}^2x \, \sigma^2 - \ln\big(\det\big((\partial_0 + \mu) \gamma_0 + \partial_1 \gamma_1 + \sigma\big)\big)\bigg)\bigg)}_{S_\textrm{eff}}}.
\end{align}
The effective action has a discrete chiral symmetry, $S_\textrm{eff}[+\sigma] = S_\textrm{eff}[-\sigma]$, where $\langle \sigma \rangle \propto \langle \sum_n \bar{\psi}_n \psi_n \rangle$ represents the chiral condensate and indicates, whether the symmetry is spontaneously broken.

\begin{figure}[htb]
\centering
\includegraphics[width=7cm]{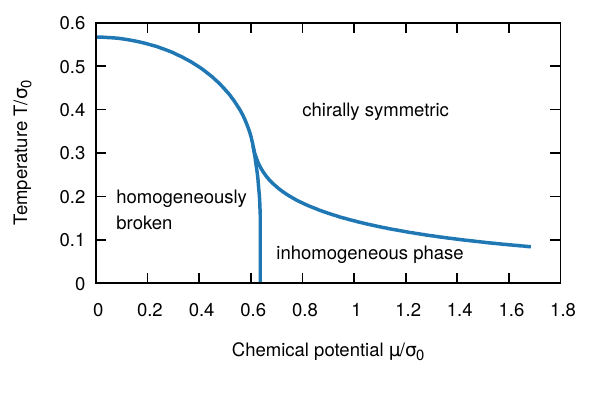}
\caption{\label{Fig:anaphase}Phase diagram of the GN model in the large-$\Nf$ limit (see \cite{Thies:2003kk,Schnetz:2004vr}).}
\end{figure}

The phase diagram of the 1+1 dimensional GN model in the limit \mbox{$\Nf \rightarrow \infty$} has been calculated in \cite{Thies:2003kk,Schnetz:2004vr}. There are three phases (see Fig.\ \ref{Fig:anaphase}):
\begin{itemize}
\item a \textit{chirally symmetric phase}, where $\langle \sigma \rangle = 0$;
\item a \textit{homogeneously broken phase}, where $\langle \sigma \rangle = \text{const} \neq 0$;
\item an \textit{inhomogeneous phase}, where $\langle \sigma \rangle$ is a spatially oscillating function.
\end{itemize}
In the inhomogeneous phase $\langle \sigma \rangle$ exhibits a periodic kink-antikink structure close to the phase boundary to the homogeneously broken phase and gradually changes into a $\sin$-like structure for increasing $\mu$.

\section{The phase diagram at finite number of fermion flavors}

We perform lattice Monte Carlo simulations of the 1+1 dimensional GN model defined in eq.\ (\ref{EQN001}) at finite $\Nf \in \{ 8 , 16 , 24 , 32 , 48 \}$. We use two different discretizations of the fermionic determinant, naive fermions and SLAC fermions (see e.g.\ \cite{Wozar:2011gu}), which we consider to be an important cross check of our numerical results: the results obtained with the two discretizations agree within statiscal errors. We set the scale via the absolute value of the chiral condensate at chemical potential $\mu = 0$ and temperature $T = 0$, i.e.\ $\sigma_0 = \langle |\bar{\sigma} | \rangle_{\mu=0,T=0}$, where
\begin{align}
\bar{\sigma} = \frac{1}{V} \sum_{x,t} \sigma(x,t),
\end{align}
$V$ is the number of lattice sites and $\langle \ldots \rangle_{\mu,T}$ denotes the path integral expectation value at chemical potential $\mu$ and at temperature $T$, i.e.\ the average over the generated set of Monte Carlo field configurations. In other words, we express dimensionful quantities in units of $\sigma_0$, e.g.\ $\mu / \sigma_0$, $T / \sigma_0$.

\begin{figure}[htb]
\centering
\includegraphics[width=6cm]{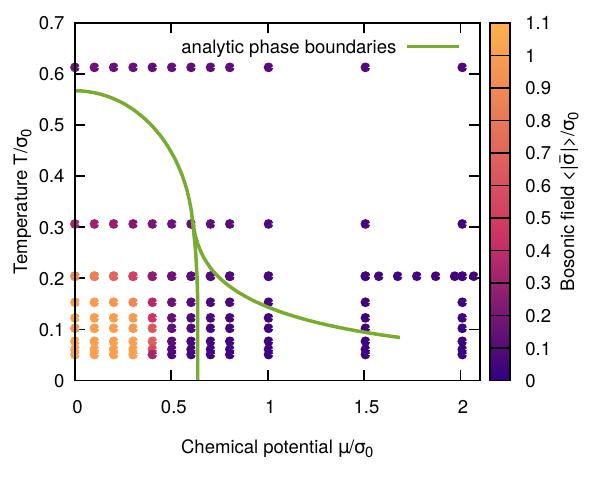}
\includegraphics[width=6cm]{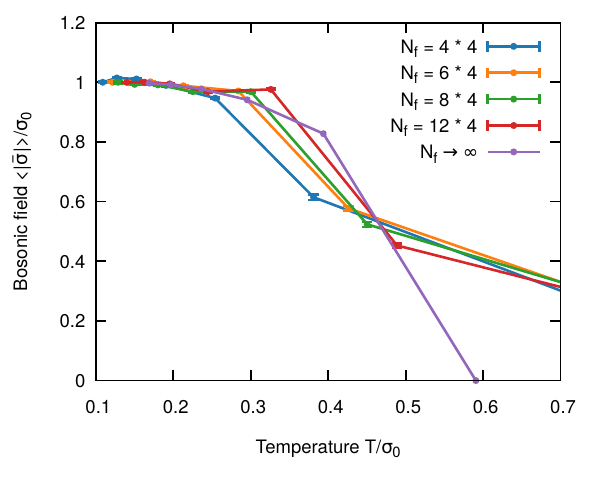}
\caption{\label{Fig:phase1}\textbf{(left)}~$\langle |\bar{\sigma} | \rangle_{\mu,T} / \sigma_0$ as a function of $\mu/\sigma_0$ and $T/\sigma_0$ for $\Nf = 8$. \textbf{(right)}~$\langle |\bar{\sigma} | \rangle_{\mu=0,T} / \sigma_0$ as a function of $T/\sigma_0$ and $\mu/\sigma_0=0$ for various $\Nf$. }
\end{figure}

$\langle |\bar{\sigma} | \rangle_{\mu,T}$ is also a suitable approximate order parameter to distinguish between a homogeneously broken phase on the one hand ($\langle |\bar{\sigma} | \rangle_{\mu,T} \neq 0$) and a restored or inhomogeneous phase on the other hand ($\langle |\bar{\sigma} | \rangle_{\mu,T} \approx 0$). Numerical results for $\Nf = 8$ are shown in Fig.\ \ref{Fig:phase1}, left plot. A homogeneously broken phase is indicated by the yellow dots at small $\mu$ and small $T$, somewhat smaller, but in a similar region as for infinite $\Nf$. Results from analogous computations for $\Nf \in \{ 16, 24, 32, 48 \}$ restricted to $\mu = 0$ are shown in Fig.\ \ref{Fig:phase1}, right plot. When increasing $\Nf$, the results approach the numerical result at infinite $\Nf$ (the latter has been obtained using techniques developed and explained in \cite{deForcrand:2006zz,Wagner:2007he,Heinz:2015lua}).

\begin{figure}[htb]
\centering
\includegraphics[width=6cm]{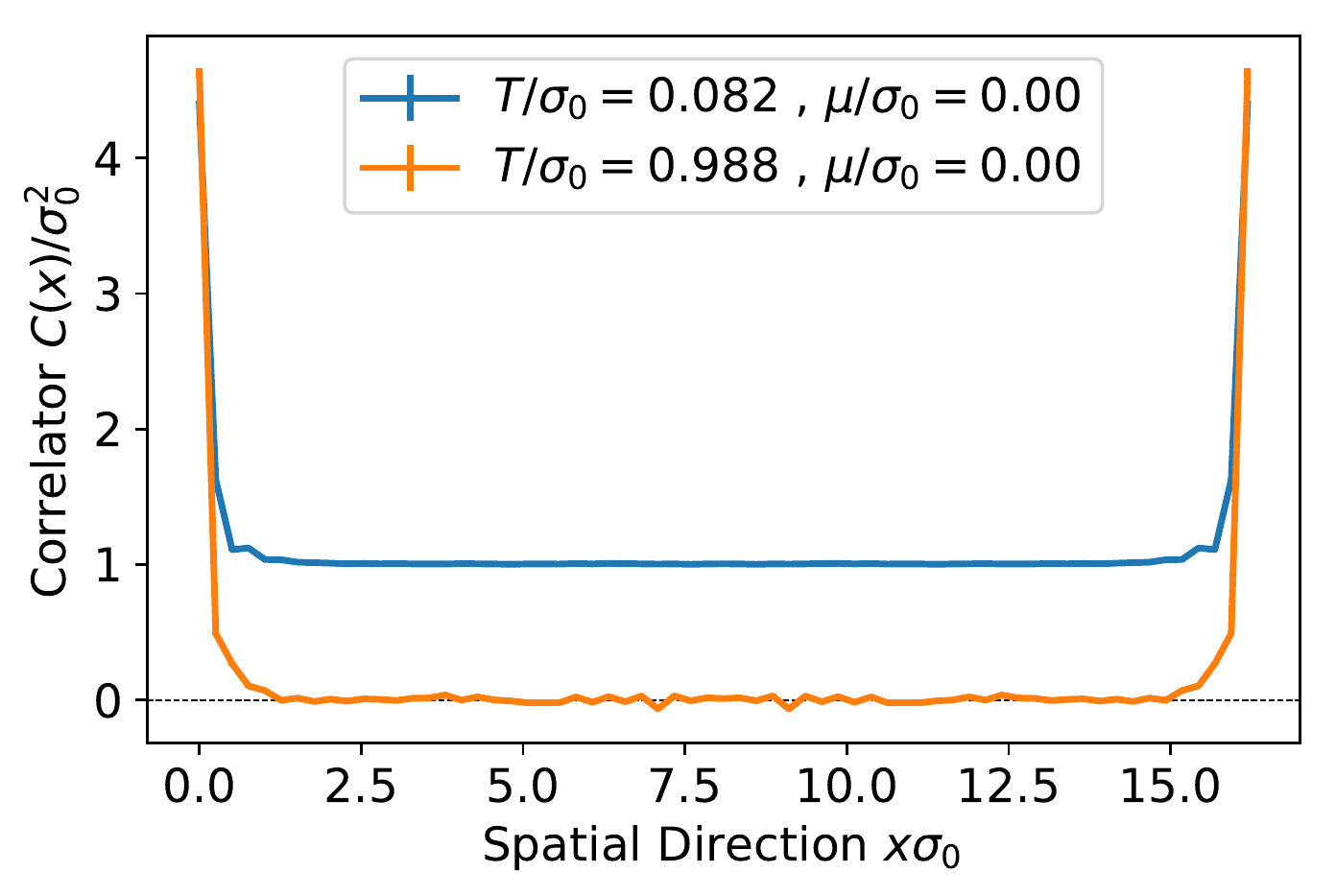}
\includegraphics[width=6cm]{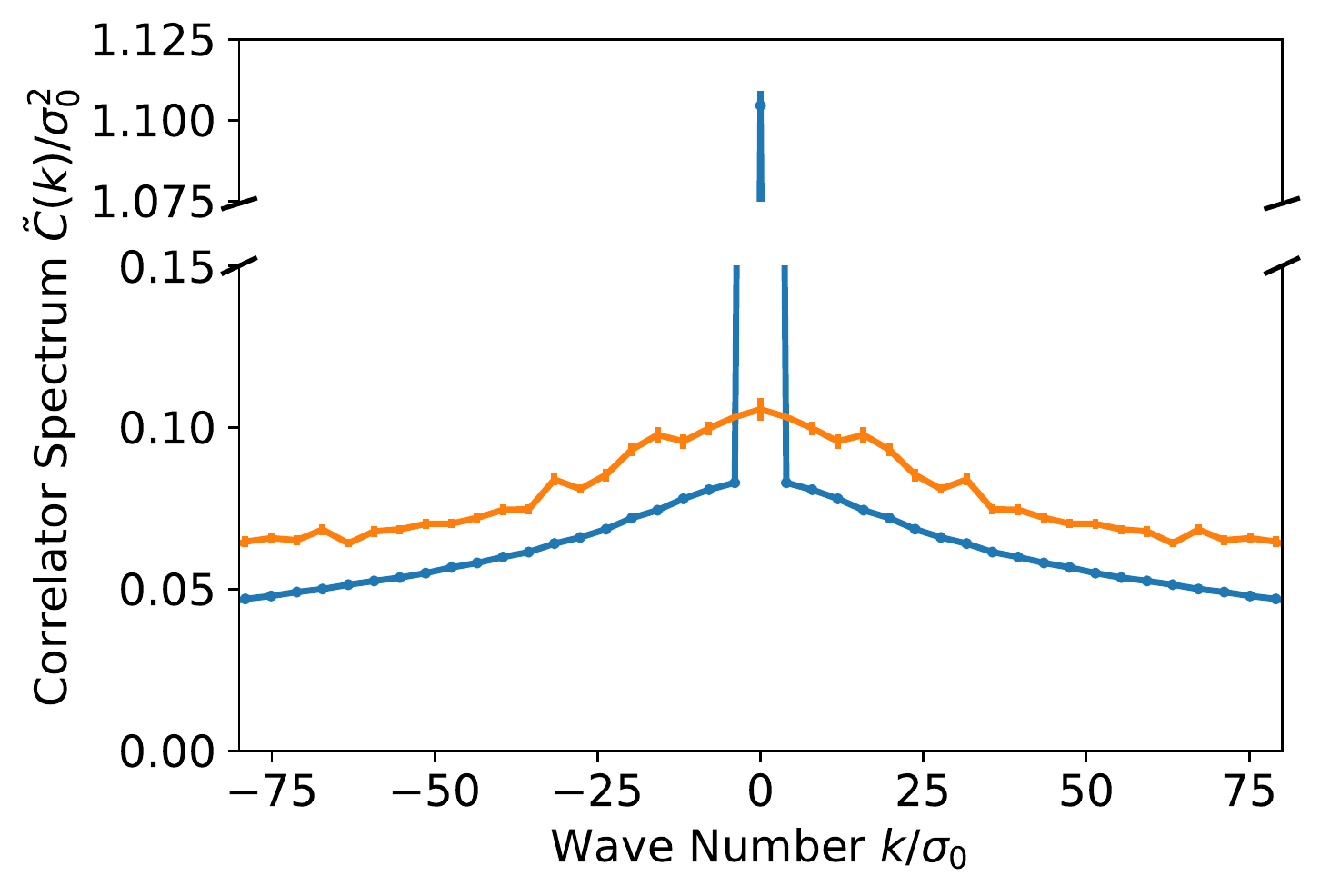}
\includegraphics[width=6cm]{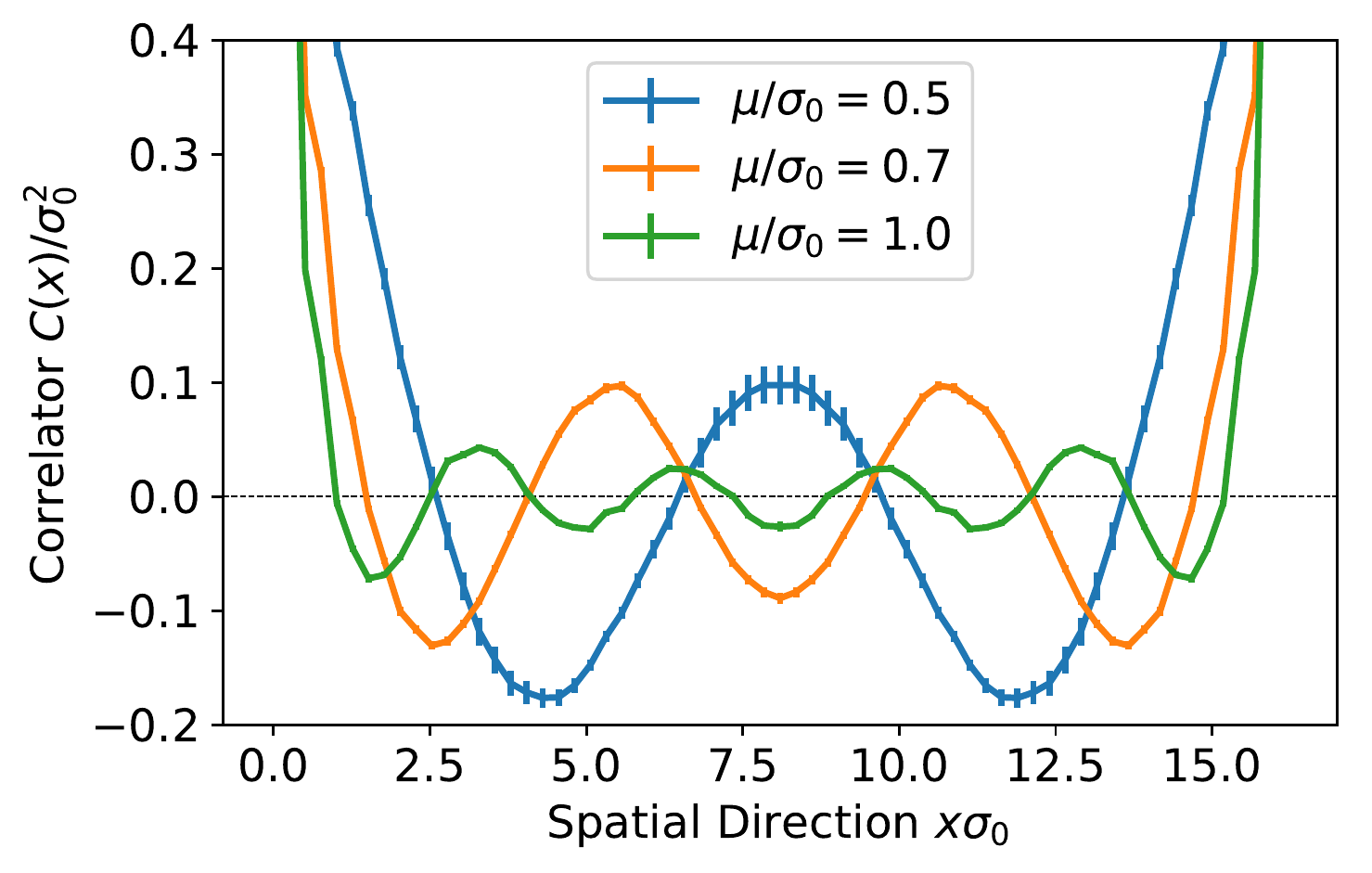}
\includegraphics[width=6cm]{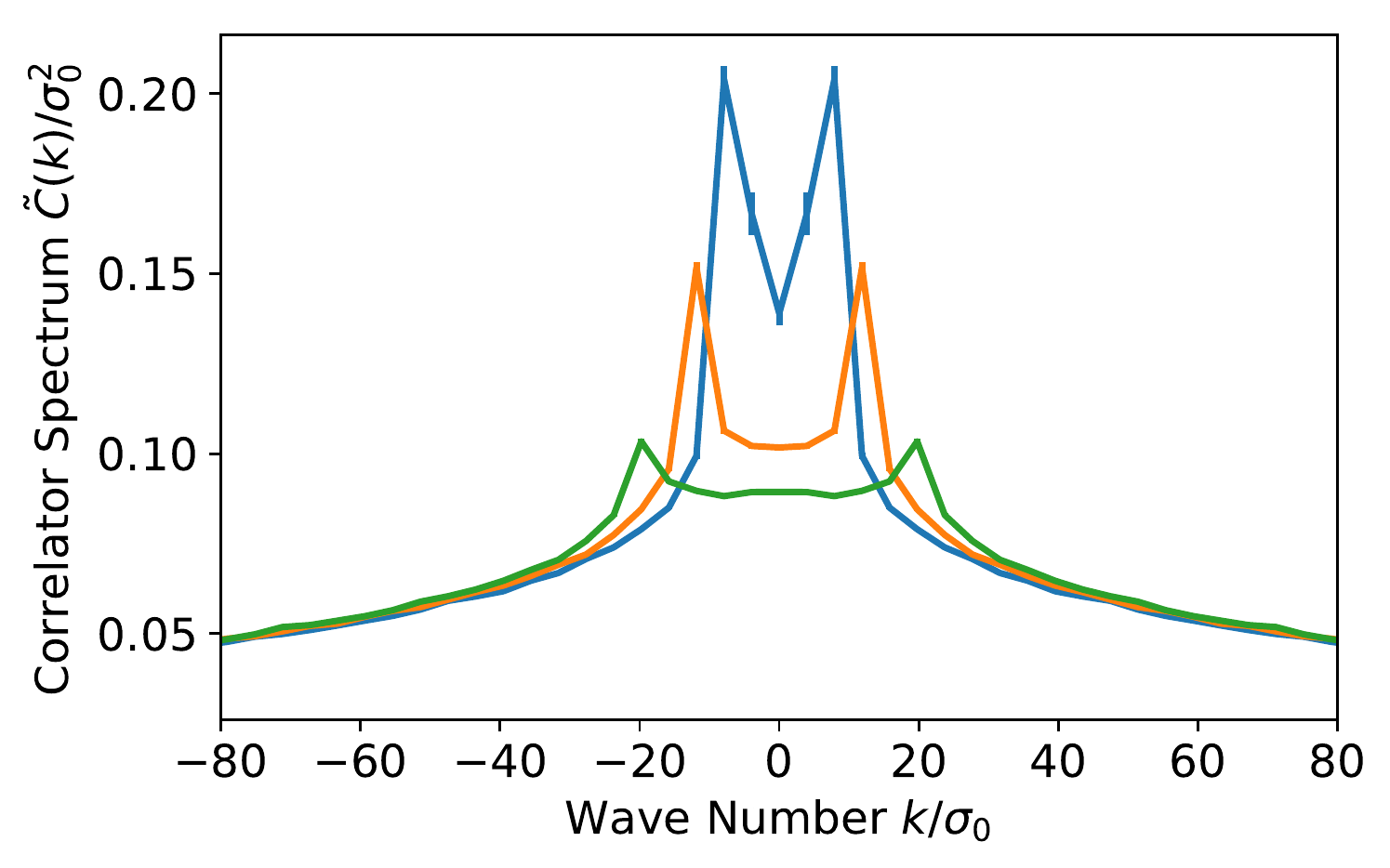}
\caption{\label{Fig:C}$C(x)$ and $\tilde{C}(k)$ for $\Nf = 8$. \textbf{(top)}~$\mu/\sigma_0 = 0$ and $T/\sigma_0 = 0.988$ (chirally symmetric phase) as well as $T/\sigma_0 = 0.082$ (homogeneously broken phase). \textbf{(bottom)}~$\mu/\sigma_0 \in \{ 0.5 , 0.7 , 1.0 \}$ and $T/\sigma_0 = 0.082$ (inhomogeneous phase).}
\end{figure}

To check for the existence of an inhomogeneous phase at finite $\Nf$, we compute the spatial correlation function of the chiral condensate $\langle C(x) \rangle_{\mu,T}$ and its Fourier transform $\langle \tilde{C}(k) \rangle_{\mu,T}$, where
\begin{align}
C(x) = \frac{1}{V} \sum_{y,t} \sum_x \sigma(y,t) \sigma(y+x,t) .
\end{align}
Both $\langle C(x) \rangle_{\mu,T}$ and $\langle \tilde{C}(k) \rangle_{\mu,T}$ are suited to distinguish the three phases we are expecting as illustrated by Fig.\ \ref{Fig:C}:
\begin{itemize}
\item \textit{Chirally symmetric phase:} $\langle C(x) \rangle_{\mu,T}$ quickly approaches $0.0$. The Fourier transform is a smooth function close to $0.0$ indicating a vanishing chiral condensate.

\item \textit{Homogeneously broken phase:} $\langle C(x) \rangle_{\mu,T}$ quickly approaches $\sigma_0^2$. The Fourier transform exhibits a pronounced peak at $k = 0$ representing the non-vanishing constant chiral condensate.

\item \textit{Inhomogeneous phase:} $\langle C(x) \rangle_{\mu,T}$ is an oscillating function. The Fourier transform exhibits a pronounced peak at $k \neq 0$ proportional to the inverse wave length of the chiral condensate.
\end{itemize}
Of particular interest are the plots at the bottom of Fig.\ \ref{Fig:C}, because they provide clear evidence for the existence of an inhomogeneous phase at finite $\Nf$.

\begin{figure}[htb]
\centering
\includegraphics[width=12.5cm]{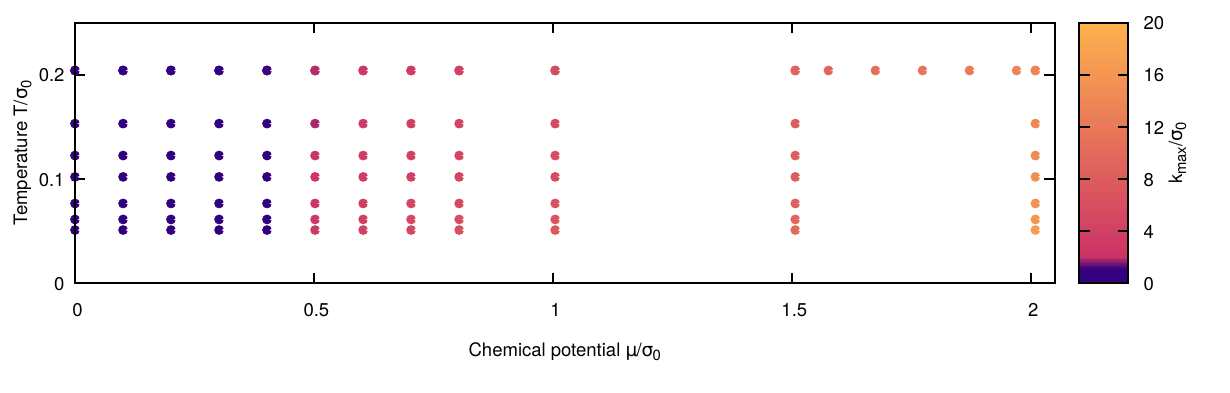}
\caption{\label{Fig:kmax}$k_\text{max}/\sigma_0$ as a function of $\mu/\sigma_0$ and $T/\sigma_0$ for $\Nf = 8$.}
\end{figure}

To identify the boundary between the homogeneously broken phase and the inhomogeneous phase, we plot in Fig.\ \ref{Fig:kmax}
\begin{align}
k_\text{max} = \Big|\arg \max \Big(\langle \tilde{C}(k) \rangle_{\mu,T}\Big)\Big|
\end{align}
as a function of $\mu$ and $T$. The phase boundary is clearly visible at $\approx \mu/\sigma_0 \approx 0.45$ separating the blue points ($k_\text{max} \approx 0$, homogeneously broken phase) from the red points ($k_\text{max} \neq 0$, inhomogeneous phase).

\begin{figure}[htb]
\centering
\includegraphics[width=4.0cm]{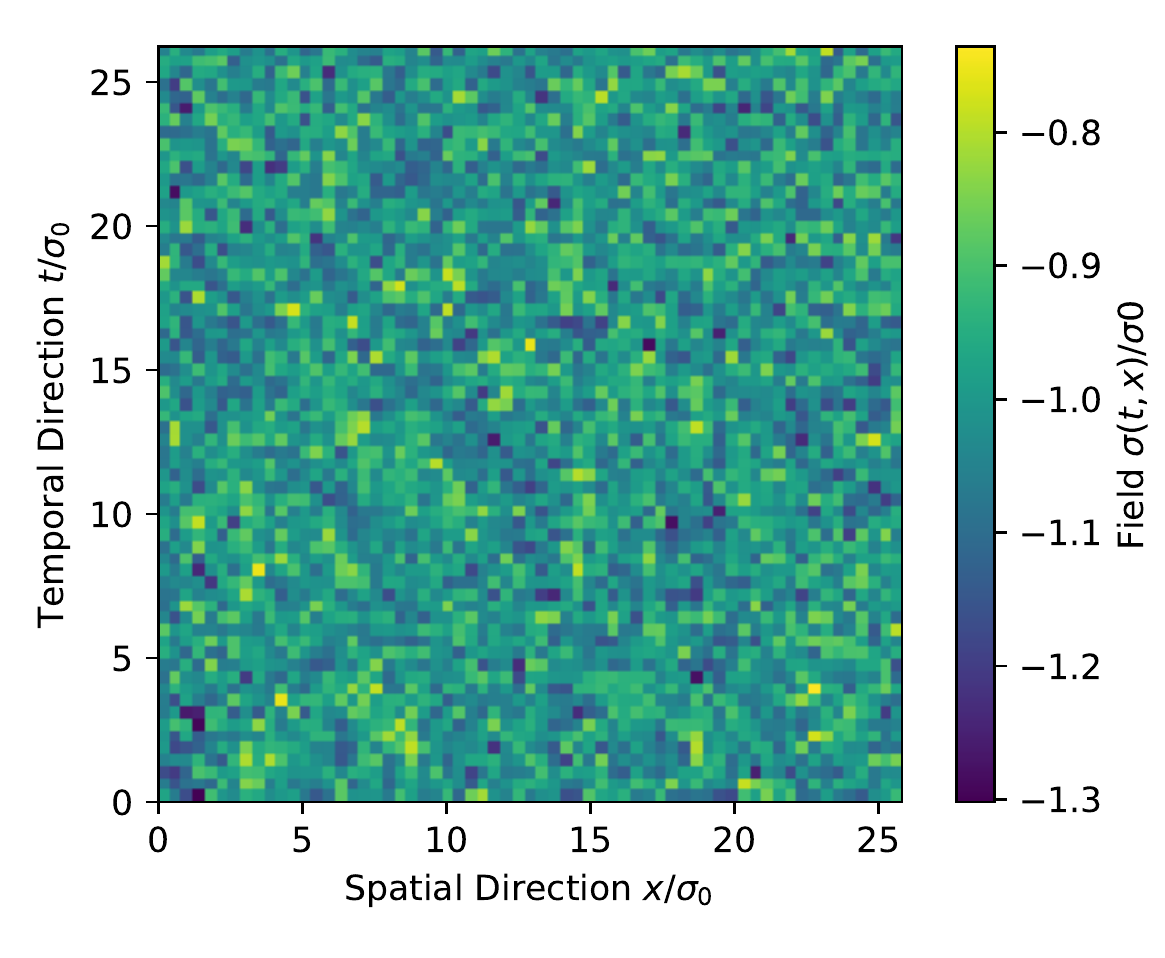}
\includegraphics[width=4.0cm]{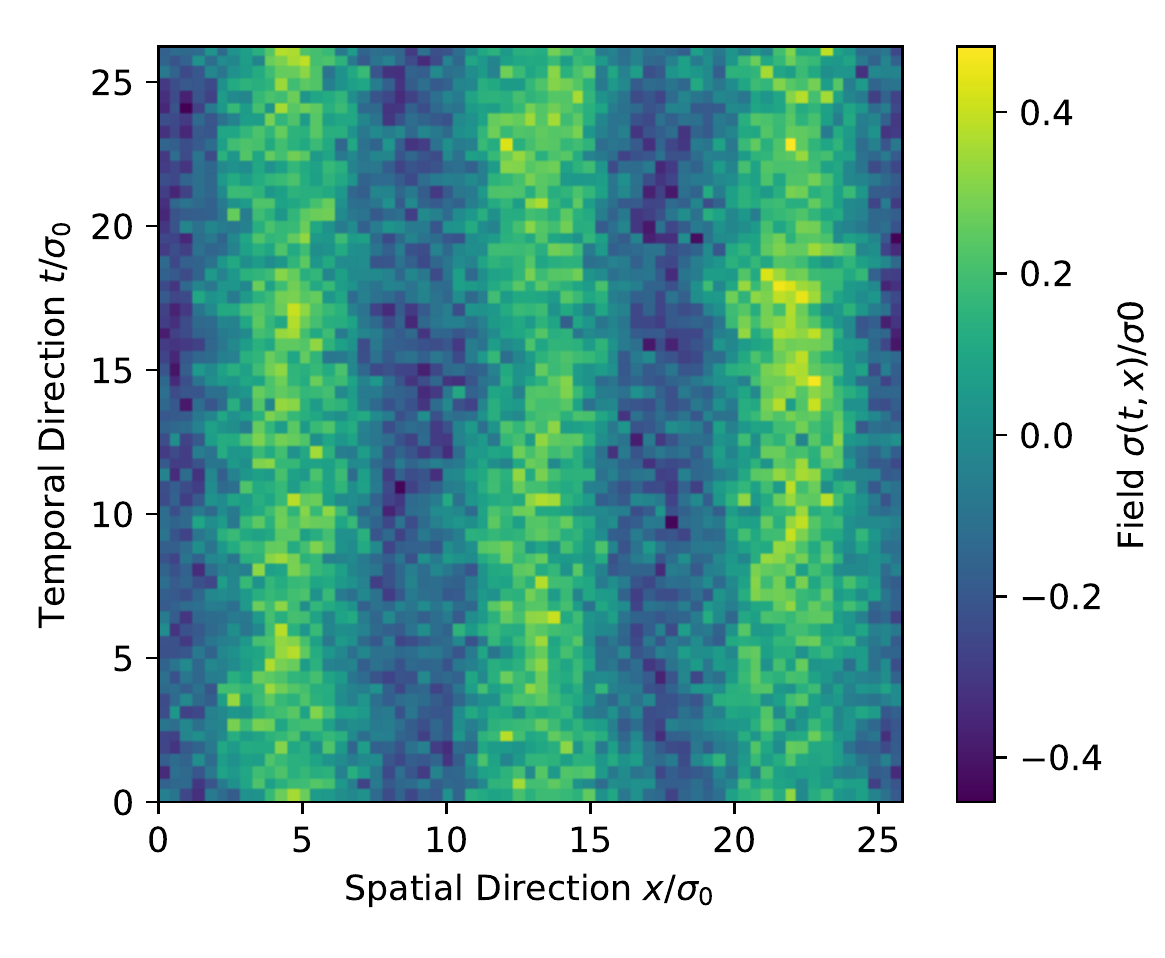}
\includegraphics[width=4.0cm]{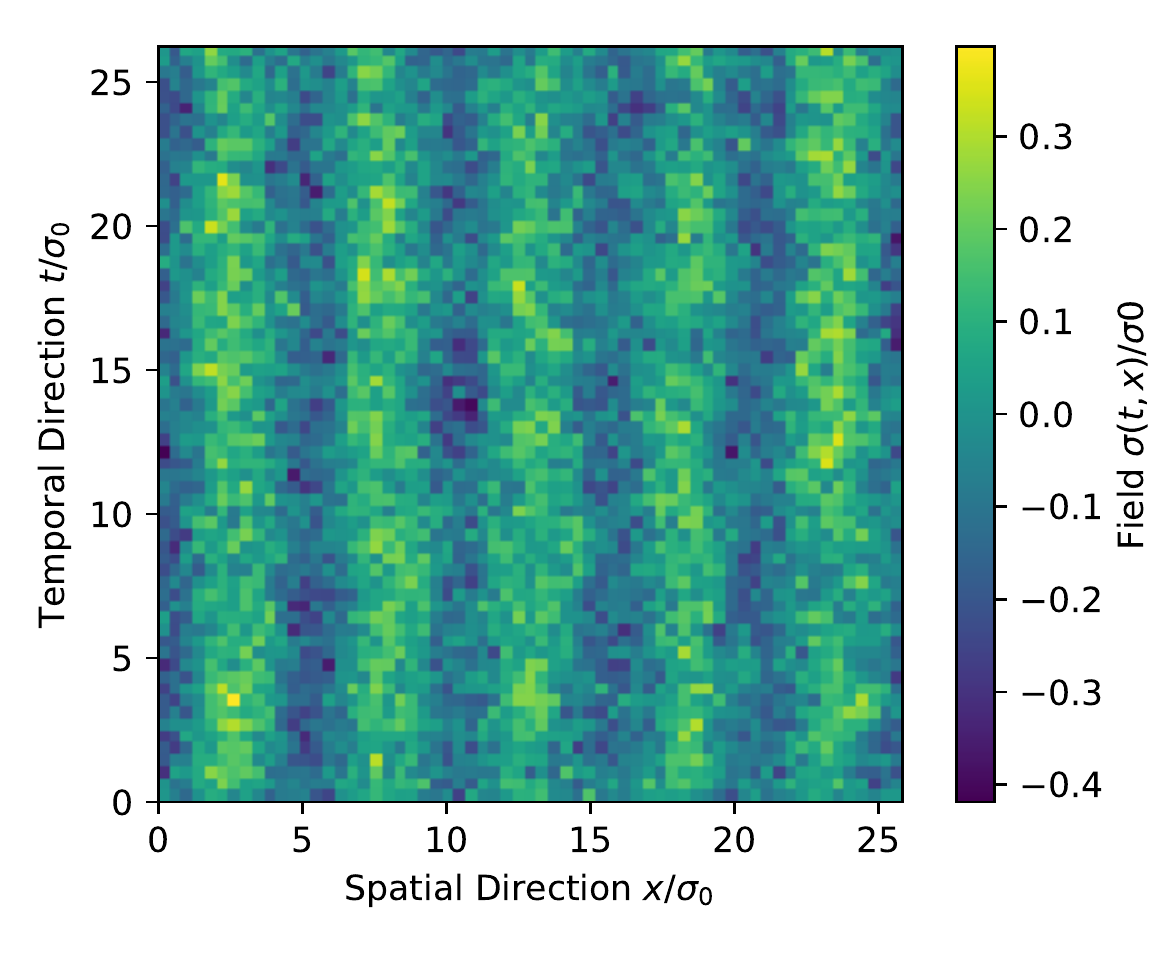}
\caption{\label{Fig:shifted}$\langle \sigma(x + \xshift,t) \rangle_{\mu,T}$ as a function of $x/\sigma_0$ and $t/\sigma_0$ for $(\mu/\sigma_0,T/\sigma_0) = (0.0,0.038)$ (homogeneously broken phase, left plot) and $(\mu/\sigma_0,T/\sigma_0) \in \{ (0.5,0.038) , (0.7,0.038) \}$ (inhomogeneous phase, center and right plot). }
\end{figure}

To exhibit the oscillations of the chiral condensate in the inhomogeneous phase in an even more direct way, we compute $\langle \sigma(x + \xshift,t) \rangle_{\mu,T}$. Here $\xshift$ is the phase shift of the spatially oscillating chiral condensate $\sigma(x,t)$ determined individually for each Monte Carlo field configuration by a standard Fourier transform. In this way destructive interference is excluded, when averaging over the Monte Carlo field configurations. In Fig.\ \ref{Fig:shifted} we show $\langle \sigma(x + \xshift,t) \rangle_{\mu,T}$ at three different $(\mu,T)$. In the left plot (homogeneously broken phase) $\langle \sigma(x + \xshift,t) \rangle_{\mu,T}$ is almost constant, close to $\sigma_0$, while in the center plot and the right plot (inhomogeneous phase) spatial oscillations are clearly visible.

\section*{Acknowledgements}

We thank M.\ P.\ Lombardo for very helpful discussions and suggestions.
We thank M.\ Ammon and F.\ Karbstein for interesting discussions
on variations of the Goldstone theorem.
We thank A.\ Königsstein and M.\ Steil for critical comments on this manuscript.
L.P.\ thanks the organizers of the ``Excited QCD 2019'' conference for the possibility to give this talk.
M.W.\ acknowledges support by the Heisenberg Programme of the DFG (German Research Foundation), grant WA 3000/3-1.
L.P.\ and M.W.\ acknowledge support by the Deutsche Forschungsgemeinschaft (DFG, German Research Foundation) through the CRC-TR 211 ``Strong-interaction matter under extreme conditions'' -- project number 315477589 -- TRR 211.
This work was supported in part by the Helmholtz International Center for FAIR within the framework of the LOEWE program launched by the State of Hesse. 
Calculations on the FUCHS-CSC high-performance computer of the Frankfurt University and the local HPC cluster of the FS-University Jena were conducted for this research. We would like to thank HPC-Hessen, funded by the State Ministry of Higher Education, Research and the Arts, for programming advice.



\end{document}